\documentclass{article}
\pdfoutput=1
\usepackage[english]{babel}
\usepackage[margin=1.3in]{geometry}
\usepackage{multicol}
\usepackage{background}
\usepackage{caption}
\usepackage{upgreek}

\SetBgScale{4}
\SetBgColor{gray}
\SetBgAngle{90}
\SetBgContents{}
\SetBgPosition{-1.5,-10}
\begin{document}

{\centering
  
  {\bfseries\Large Integration of acoustical sensors into the KM3NeT Optical Modules\bigskip}
  
  A.~Enzenh\"ofer on behalf of the KM3NeT Collaboration \\
  {\it{Friedrich-Alexander-Universit\"at Erlangen-N\"urnberg, Erlangen Centre for Astroparticle Physics, Erwin-Rommel-Str.\,1, D-91058 Erlangen, Germany}} \\
  \Large August 20, 2014
  
}

\begin{abstract}
  The next generation multi-cubic-kilometre water Cherenkov neutrino telescope will be build in the Mediterranean Sea.
  This telescope, called KM3NeT, is currently entering a first construction phase.
  The KM3NeT research infrastructure will comprise 690 so-called Detection Units in its final design which will be anchored to the sea bed and held upright by submerged floats.
  The positions of these Detection Units, several hundred metres in length, and their attached Optical Modules for the detection of Cherenkov light have to be monitored continously to provide the telescope with its desired pointing precision.
  A standard way to do this is the utilisation of an acoustic positioning system using emitters at fixed positions and receivers distributed along the Detection Units.
  The KM3NeT neutrino telescope comprises a custom-made acoustic positioning system with newly designed emitters attached to the anchors of the Detection Units and custom-designed receivers attached to the Detection Units.
  This article describes an approach for a receiver and its performance.
  The proposed Opto-Acoustical Modules combine the optical sensors for the telescope with the acoustical sensors necessary for the positioning of the module itself.
  This combination leads to a compact design suited for an easy deployment of the numerous Detection Units.
  Furthermore, the instrumented volume can be used for scientific analyses such as marine science and acoustic particle detection.
\end{abstract}

\section{Introduction}

The KM3NeT~\cite{km3net} research infrastructure houses the next generation large scale neutrino telescope and will be located in the Mediterranean Sea.
It succeeds the ANTARES~\cite{antares} neutrino telescope also located in the Mediterranean Sea.
The KM3NeT telescope will eventually consist of 690 Detection Units (DUs).
Each DU will be anchored to the sea bed and held upright by submerged floats.
About 18 Digital Optical Modules (DOMs) built in the KM3NeT multi-PMT design~\cite{km3net_tdr} are distributed at equal distances along each DU.
The movement of the upper parts of the DUs caused by sea currents makes it mandatory to monitor in principle each DOM's position within the detector.
This monitoring is achieved by an acoustic positioning system (APS) which will be described later on.
This article describes the implementation of a dedicated APS~\cite{aps}, especially with a special solution for the acoustic receivers.
These receivers will be installed inside the DUs, simplifying the design of the Optical Modules and the deployment of the DUs as compared to a design with external acoustical sensors.
A key consideration for the design is to avoid any interference with the optical part of the detector.
It is possible to extend the main purpose of position calibration to further tasks such as oceanography or the detection of acoustic signals produced by neutrinos as described in the thermoacoustic model~\cite{thermo_acoustic_model}.
The ANTARES neutrino telescope provides the framework for tests under real conditions.
The ANTARES neutrino telescope also houses the AMADEUS setup~\cite{amadeus} which was designed to investigate the feasibility of the acoustic particle detection technique.
Investigations in the context of this feasibility study led to the current design of an ``Opto-Acoustical Module'' (OAM, cf.~Fig.~\ref{fig:lom} center) presented in this article.

\section{Acoustic positioning system}

An acoustic positioning system in general consists of emitters at well-known positions and receivers fixed to the structure for which the position needs to be determined.
By taking advantage of the rigidity of the DU structure and by applying a realistic model of the DU shape it is possible to achieve a position accuracy of better than $ 40\,\textrm{cm} $ as required in~\cite{km3net_tdr} with a reduced amount of acoustic receivers, i.e.~not every single DOM has to be monitored individually.
Each emitter emits a unique signal through synchronised commands in order to identify the corresponding emitter and the emission time as required for the calculation of the travel time of the acoustic signal from the emitter to the receiver.
Additional sensors measure the local speed of sound, which depends on several environmental conditions such as sea temperature, salinity and ambient pressure.
Together with this information, the relative position of the receiver to the emitters can be obtained through trilateration given a set of emitters:
\begin{equation}
  \label{equ_1}
  \left|\vec{r}_{\textrm{\small{reception}}} - \vec{r}_{\textrm{\small{emission}}} \right| = c_{\textrm{\small{sound}}} \times \left( t_{\textrm{\small{reception}}} - t_{\textrm{\small{emission}}} \right) \quad .
\end{equation}
In this way it is possible to determine the position of the receiver with sufficient accuracy.
For the KM3NeT neutrino telescope, a custom-made system was chosen over a commercial system.
The reason for this is to increase the performance of the APS in terms of e.g.~utilisable acoustic signals and number of receivers and emitters.
Typical emission sensitivities of the emitters are of the order of $ 140\,\textrm{dB re}\,1\,\upmu \textrm{Pa}/\textrm{V}\,@\,1\,\textrm{m} $.
This emissivity is common to most devices as it strongly depends on the geometry of the emitter as well as its electronics.
There is only a narrow margin to vary this quantity.
There are ample options in case of the emitted signals by contrast.
Robust signals with unique features can be detected even in varying ambient background conditions with sufficient accuracy.
These signals can be composed of signals with orthogonal bases or some other sort of robust signals.
The receivers have to be adapted to these different signals.
Two types of receivers will be described in the next section.

\section{Hydrophones and Opto-Acoustical Modules}

\begin{figure}
  \includegraphics[height=.22\textheight]{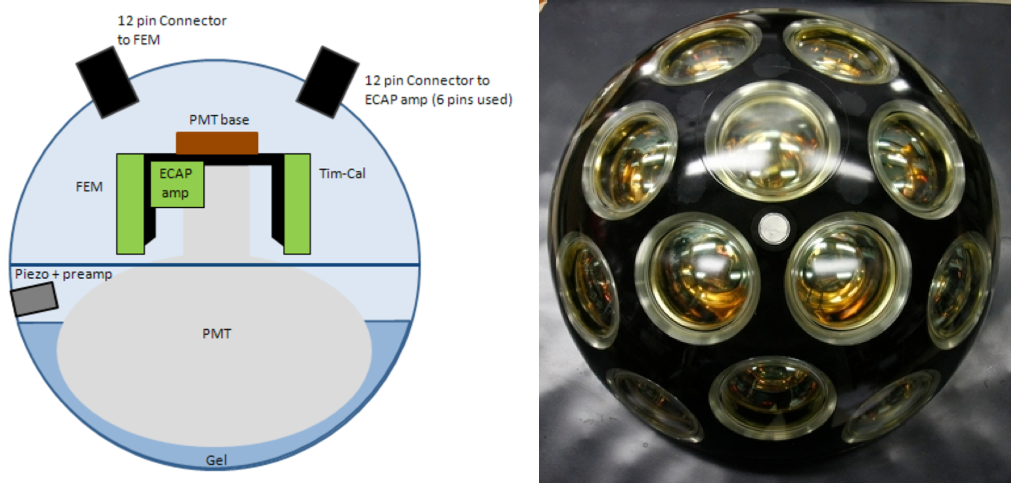}
  \includegraphics[height=.22\textheight]{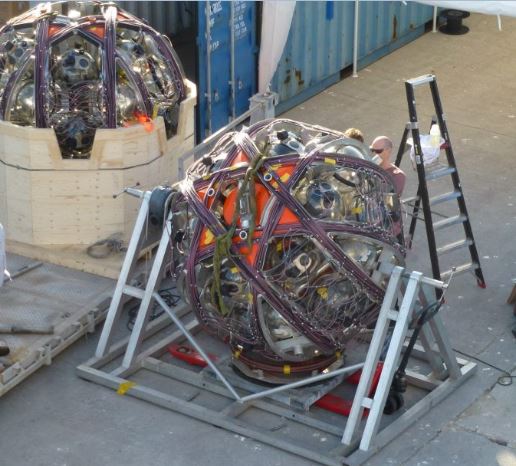}
  \caption{{\it{Left:}} This schematic depicts the realisation of an OAM upgrading a single PMT optical module as used for the NEMO Phase \MakeUppercase{\romannumeral 2} optical module. The acoustical sensor is located near the equator on the left side (labelled ``Piezo + preamp''). {\it{Center:}} The multi-PMT design as it will be used in the KM3NeT neutrino telescope. The densely packed PMTs are equally distributed inside the sphere. The small space left between the PMTs is perfectly suited for the small acoustical sensor (argentic spot in the center of the sphere). {\it{Right:}} The prototype of a Launcher of Optical Modules (LOM) during last checks before deployment. On the left another LOM in its carrier box is shown, demonstrating the space-saving design.}
  \label{fig:lom}
\end{figure}

The standard devices used as acoustic receivers, piezo ceramics based sensors moulded in a protective casing, will be denoted as hydrophones.
Their casing protects the inner part of the device from environmental influences and ensures an ideal coupling of sound signals from the water to the sensor.
Typical sensitivities of hydrophones after a first internal amplification stage range from $ -200\,\textrm{dB re}\,1\,\textrm{V}/\upmu\textrm{Pa} $ to $ -140\,\textrm{dB re}\,1\,\textrm{V}/\upmu\textrm{Pa} $ depending on their exact purpose and the used infrastructure.
Hydrophones have at least one inherent major drawback although they are available in a variety of different designs, sensitivities and thus adoptable to any conceivable neutrino telescope design:
They constitute an additional mechanical structure differing from the large amount of uniform structures needed for next generation neutrino telescopes.
The deployment of several hundred Detection Units as in the case of the KM3NeT telescope necessitates efficient deployment strategies to reduce time consuming procedures and to reduce costs.
For the KM3NeT research infrastructure it is therefore planned to use the so called ``Launcher of Optical Modules''~\cite{lom} (LOM, cf.~Fig.~\ref{fig:lom} right).
The complete detection unit is spooled on this spherical device.
This leads to a compact structure and eases the transport of several LOMs during one sea campaign.
Once the LOM is released on the seabed it unfurls autonomously by buoyancy.
The unfurled LOM can be reused again.
The spooling of additional mechanical structures into this LOM is indeed feasible but houses a significant risk in the unfurling procedure.
This is a point where an integrated Opto-Acoustical Module could be used to remove any concerns.
An OAM as referred to in this article is a standard DOM as used throughout the detector with an integrated acoustical sensor.
This acoustical sensor is also based on a piezo ceramic equipped with adequate electronics.
The sensor is directly glued to the inside of the DOM.
This procedure requires no protective casing of the sensor against environmental influences and thus allows for a very compact and inexpensive design of the acoustical sensor.
This compact design also allows for another realisation of the OAM which is currently in operation inside the NEMO~\cite{nemo} Phase \MakeUppercase{\romannumeral 2} optical module (cf.~schematic in Fig.~\ref{fig:lom} left).
This implementation will not be discussed in detail in this article but is mentioned here as it demonstrates the feasibility of the concept.
The mismatch in the acoustic impedance between sea water, glass and piezo ceramic is not crucial for the determination of the arrival times of acoustic signals but may influence advanced analyses of signal features.

\section{Performance of Opto-Acoustical Modules}

\begin{figure}
  \includegraphics[height=.3\textheight]{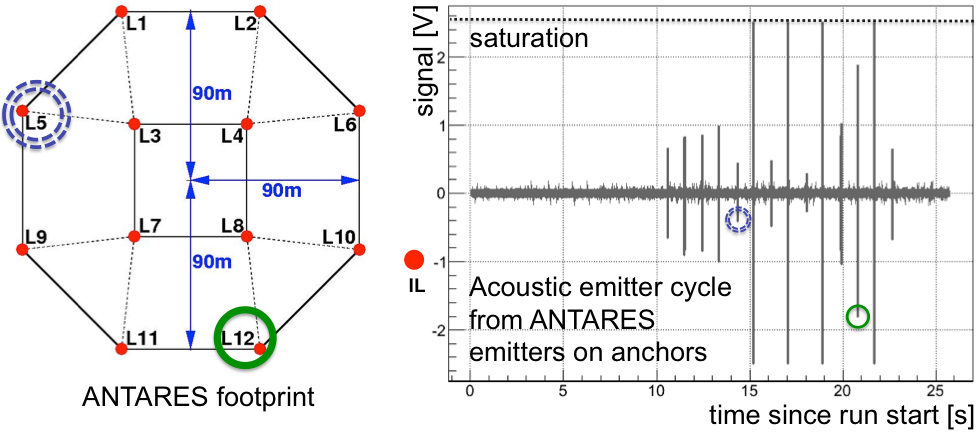}
  \caption{{\it{Left:}} Footprint of the ANTARES detector showing typical distances between the lines as well as the naming scheme of the different lines. L5 (dashed circle) and L12 (full circle) show two lines with different distances to the OAM located on the IL. This distance correlation is clearly visible on the right. {\it{Right:}} An example for an acoustic positioning cycle. Emitters at the anchors of the 12 lines and the IL send signals in chronological order, followed by a $ 14^{\textrm{\tiny{th}}} $ emission of line 2. These signals are subsequently recorded by the acoustical sensor of the OAM. All signals are clearly identifiable above the background noise. The signals saturating the acoustical sensor constitute no principal problem to the determination of the signal arrival time. The sensitivity of the acoustical sensor has been adjusted accordingly for the next iteration.}
  \label{fig:footprint}
\end{figure}

A field test in the framework of the ANTARES detector is ongoing to compare the performance of both OAM and hydrophone.
The operation of the acoustical sensor inside the OAM has to proof the ability to reconstruct the module position to a sufficient precision if it should be used as a substitute for hydrophones.
The ``Instrumentation Line'' (IL) is the additional $ 13^{\textrm{\tiny{th}}} $ line of the ANTARES detector which houses environmental monitoring devices as well as a part of the AMADEUS test setup.
After a maintenance operation on the IL, it was possible to equip one of its storeys with both OAM and hydrophone thus enabling a direct comparison of both devices.
The deployment and subsequent connection in this reconfigured configuration took place in April 2013.
Figure~{\ref{fig:footprint}} shows the ANTARES footprint with the positions of all lines.
At the anchor of each line an acoustic emitter is installed which sends acoustic signals with predefined frequencies and time intervals.
The diagram in figure~\ref{fig:footprint} shows a sample measurement of such an emission cycle and demonstrates the possibility of the OAM to detect all 12 detection lines of the ANTARES neutrino telescope.
Some signals are saturating the acoustical sensor of the OAM and some signals are very low compared to the other signals.
This effect can be attributed mainly to the angular acceptance of the acoustical sensor inside the OAM.
Other effects include the different travel distances as well as some shielding effects due to the lines in between emitter and receiver.
Nevertheless, all signals are clearly visible above the background.

\begin{figure}
  \includegraphics[width=.5\textwidth]{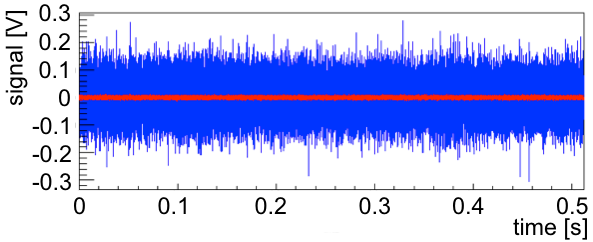}
  \includegraphics[width=.5\textwidth]{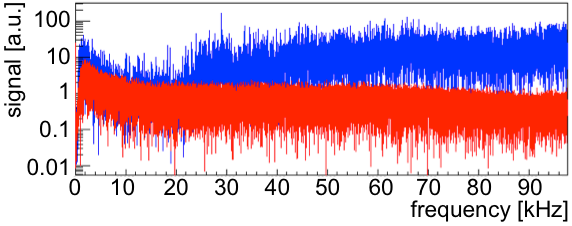}
  \caption{{\it{Left:}} A time domain background sample. The red (pale) signal corresponds to the intrinsic noise of the acoustical sensor. The blue (dark) signal shows the significant increase of the noise level after the PMTs inside the OAM are switched on.
    {\it{Right:}} The signals on the left transferred into the frequency domain. The red (pale) spectrum shows the intrinsic flat amplitude response of the acoustical sensor for frequencies above $ 10\,\textrm{kHz} $. The blue (dark) spectrum shows that the increase of the noise level is strongest in the frequency range above $ 25\,\textrm{kHz} $.}
  \label{fig:noise}
\end{figure}

However some drawback arise due to the compact design.
A KM3NeT DOM houses 31 PMTs inside a glass sphere.
In this densely packed environment the high voltage supply for the PMTs generate an electromagnetically ``loud'' environment leading to an increased noise level for the acoustical sensor.
This behaviour can be seen in figure~{\ref{fig:noise}}.
The determination of the rather strong and well defined signals of the acoustic positioning system is not crucial affected by this.
Advanced algorithms can significantly increase the detectability of well defined signals.
This increased noise level could cause difficulties if the data should be used for additional purposes like acoustical particle detection or marine science because in these cases, a variety of different signals and signal attributes have to be extracted from the measured spectra.

\section{Conclusions and outlook}

The feasibility of the construction and operation of OAMs in the KM3NeT multi-PMT design was demonstrated within the ANTARES framework.
Its use for the position determination of the module is confirmed even with the increased noise level during operation due to the high voltage supply of the PMTs.
However, the acoustical sensor will be redesigned for the final use in the KM3NeT neutrino telescope.
The analogue signal path will be reduced to a minimum in order to reduce the electromagnetic interference.
In addition the PMT bases have been redesigned to minimise emission of electromagnetic noise.

\section*{Acknowledgments}

The research leading to these results in parts has received funding from the European Community Seventh Framework Programme under grant agreement n.~212525 and from the Sixth Framework Programme under contract n.~011937.

\end{document}